\documentclass[twocolumn]{IEEEtran}
\usepackage{booktabs}
\usepackage{url}
\usepackage{caption}
\usepackage{graphicx}
\usepackage{array}
\usepackage{tabularx}
\usepackage{tabu}
\usepackage{cite}

\begin{document}

\title{Persuasive Technology For Human Development: Review and Case Study}
\author{Ali Harris, Saif ul Islam, Junaid Qadir, Ussama Ahmad Khan\\
Information Technology University (ITU), Punjab, Lahore, Pakistan\\
mscs15031@itu.edu.pk, mscs13053@itu.edu.pk, junaid.qadir@itu.edu.pk, ussama.ahmad@itu.edu.pk}

\maketitle

\begin{abstract}
Technology is an extremely potent tool that can be leveraged for human development and social good. Owing to the great importance of environment and human psychology in driving human behavior, and the ubiquity of technology in modern life, there is a need to leverage the insights and capabilities of both fields together for nudging people towards a behavior that is optimal in some sense (personal or social). In this regard, the field of persuasive technology, which proposes to infuse technology with appropriate design and incentives using insights from psychology, behavioral economics, and human-computer interaction holds a lot of promise. Whilst persuasive technology is already being developed and is at play in many commercial applications, it can have the great social impact in the field of Information and Communication Technology for Development (ICTD) which uses Information and Communication Technology (ICT) for human developmental ends such as education and health. In this paper we will explore what persuasive technology is and how it can be used for the ends of human development. To develop the ideas in a concrete setting, we present a case study outlining how persuasive technology can be used for human development in Pakistan, a developing South Asian country, that suffers from many of the problems that plague typical developing country.

\end{abstract}

\section{Introduction} 

The change of human mind to behave positively in situations has always been the target of many researchers. The domain of Human Computer Interaction (HCI)---which leverages techniques from computer science and several other disciplines like human behavior, psychology and cognitive science---focuses on the design of technology so that human computer interaction becomes as intuitive and user friendly as possible. The interaction provides a powerful means for creating a strong bond between the two entities. Embodied agents are one of the major areas where the principles of HCI are implemented as embodied agents aim to provide users with much interactions, which is provided in the fields of health, education and learning, video games and some military applications \cite{beale2009affective}.

In this paper, we are focusing on the use of ``\textit{persuasive technology}''---which refers to technology that is designed to persuade humans by changing their attitude and influencing their minds regarding various matters (economical, medical, or personal) using technological resources \cite{fogg2002persuasive}---for human development. In particular, we are interested in using persuasive technology to influence people \textit{voluntarily} towards pro-social behaviors and away from anti-social behaviors \cite{ijsselsteijn2006persuasive}. We note that the act of persuasion is starkly different from a forceful implementation or coercion through the use of warnings or fear. The act of persuasion is also expected to follow some ethical guidelines and avoid any deception to achieve some target. Persuasive technology is broadly applicable and can be applied to commerce, education and learning, safety, environmental preservation, occupational effectiveness, preventive health-care, fitness and well-being, disease management, personal finance, community involvement, personal relationship, personal management and self improvement \cite{fogg2002persuasive}.

The rise of Facebook, Twitter and other digital social media websites is a testament to the potency of technology to persuade people to buy, sell or engage in certain behaviors. The use of persuasive technology is on the rise in multiple domains where technology can be used as an effective medium to nudge human behavior.  In this regard, a new field of study called ``\textit{Captology}'', a term coined by Stanford researcher BJ Fogg \cite{fogg2009motivating}, has been proposed which focuses on the design, research and analysis of interactive computing products created for the purpose of changing people's attitudes or behaviors. Web browsers are being used as a medium for behavior change interventions through the use of subliminal priming. For example, \textit{Subly}, a Chrome browser extension, primes behavioral concepts through emphasis on words or phrases when people are browsing the internet \cite{caraban2017design}. Similarly online tools are being developed that increase engagement by learning about energy awareness \cite{hedin2017kilowh}.Furthermore, persuasive technology can also be deployed to break undesirable habits. For example, Wood and Neal suggest a strategy known as ``vigilant monitoring'' which brings together conscious thoughts of control as well as automatic process in the brain \cite{wood2016healthy}.

%According to the author, this strategy can be most effective in order to stop or reduce undesirable behavior that occurs in day to day life. The authors also suggest ways in which technology can facilitate vigilant monitoring for individuals.  

%Computers can be designed in such a way that they behave intelligently by incorporating emotions and feelings just like human beings. This emerging field of research which involves engendering human-like feelings and emotions in computers is known as \textit{Affective Computing}. This growing field of research focuses on making the computers intelligent enough to have feelings and to behave in different scenarios according to some emotion \cite{vesterinen2001affective}. There are certain applications presented for computer-assisted learning, perceptual information retrieval, arts and entertainment, human health and interaction. To make applications of computer that can interact with human intelligently and in a natural manner, computers need to have the recognizing ability and express the emotions well \cite{picard1997affective}.%

To the best of our knowledge, this is the first study that has studied the applications of persuasive technology for human development in the context of a case study of a developing country. In this regard, we first present a general discussion of the various application areas in which persuasive technology can help modify human behavior towards positive social outcomes in developing countries, and then anchor the discussion in a concrete setting by discussing the case of Pakistan, a developing country in South Asia. Before going any further, we discuss the organization of the paper in what follows. Section II provides the background study of related fields and designing of technology. In Section III, a thorough explanation of applications that involve persuasive technology is given, Section IV elaborates case studies with context to Pakistan and how behavior change through technology can have positive impact on people. In Section V, we discuss some of the future opportunities and pitfalls in the area of persuasive technology and finally Section VI provides the conclusion of the paper.

\section{Background}

\subsection{Related Fields}

In this section, we elaborate several domains from the fields of social sciences, psychology and economics that are focused on engendering behavioral changes in humans for their welfare and for better life styles. 

%It includes nudge theory, behavioral economics from the domain of psychology and some other principles like positive computing which tend to improve human behavior for a positive change. 

%\subsubsection{Human Computer Interaction (HCI)}

\vspace{2mm}
\subsubsection{Affective Computing}

In order to build technology with human-like emotions and behavior, there is a need to build computers intelligent enough to act and behave like human beings while making decisions about their daily life routines. To achieve this purpose a new domain in computer science known as \textit{Affective Computing} can be very useful as it is aimed to empower computers with abilities similar to human behaviors \cite{vesterinen2001affective}. Computers need to have the recognizing ability and express the emotions naturally like humans \cite{picard1997affective}. There are certain applications available for computer-assisted learning, perceptual information retrieval, arts and entertainment, human health and interaction. 

\vspace{1mm}
\subsubsection{Nudge Theory/ Libertarian Paternalism}

Nudge theory focuses on developing a choice architecture so that users are inclined towards making the right or beneficial choice in a particular situation \cite{sunstein2008nudge}. According to Richard Thaler and Cass Sunstein, two behavioral economists who founded this theory, human beings are prone to make mistakes, but they can make better decisions if they are given good choices (and particularly, good defaults) \cite{leonard2008richard}. For example, according to one study, when people were provided with a comparison of their consumption of electricity with average user consumption  on their electricity bills, the one who had high electricity consumption restricted themselves to achieve a lower consumption in order to receive low cost of electricity \cite{leonard2008richard}. Another example of nudge theory at work is the website called Stikk\footnote{\url{http://www.stickk.com/}}, which helps to achieve personal goals (such as losing weight, quitting smoking, staying healthier, etc.). Once we set some goal, like to jog three times a day, we have to put some money at stake and if we fail to meet our own set target, all money would be donated to any charity organization. It also features to invite our friends who can become referee to monitor the accuracy. 

Using GPS technology and sensors in smartphones, tailored nudges can be given to individuals on the basis of their physical proximity to locations that are typically associated with unwanted habits such as fast-food restaurants or by identification of certain daily habits such as watching TV or sitting at the same location for more than what is desirable. These nudges can be in the form of simple reminders such as ``Don't Go'' or ``Order this and not that''. There is an ongoing debate on whether the nudges are a useful way to improve health sector as many of the libertarians do not believe in it, or it is hard to make them realize that nudging the behavior is a useful way to promote healthier life since they believe that many people do not know what is best for them. It also contradicts the psychology of libertarian people as they argue that nudge theory limits the freedom of choices \cite{vallgaarda2012nudge}. Sunstein offers a counter-view by arguing that choice architectures are embedded implicitly in all decisions and discusses various ethical issues related to nudge theory in detail\cite{sunstein2014ethics}, some of these issues are discussed later in paper.

\vspace{2mm}
\subsubsection{Positive Psychology and Positive Computing}

Positive Computing is defined as a field which combines human psychology, HCI, neuroscience, social sciences, behavioral economics and education to implant new methods of creating technology that can foster psychological well-being in humans. This new era of technology helps in making perfect decisions and support abilities for a positive change in people \cite{calvo2014positive}, like changing people attitudes towards goal setting to quit smoking habits or setting goals to loose weight using website like Stikk.

\vspace{2mm}
\subsubsection{Behavioral Economics}

Human minds tend to behave in a biased manner naturally which is rational thinking and in contrast to this rational behavior, there is another process of thinking called intuitive thinking process as described by Kahneman in his book ``Thinking, Fast and Slow'' \cite{kahneman2011thinking}. The behavioral economics, which is a sub domain of psychology and economics, tends to measure this quality of making rational choices in certain areas where people make choices from the available options in their own interests \cite{gennetian2011behavioral}. The aim of persuasive technology is same that using the technology we can limit user to choose best among choices available.

%The term behavioral economics is a field of study from multiple disciplines including human psychology and economics to measure the limitations of human and its complications. Behavioral economics deals with psychological effect on human and how this effect the economic actions of people in certain conditions. Behavioral economics believes that people make decisions that are biased towards a predictable manner \cite{mullainathan2000behavioral}. The impact of mind in thinking process is very critical to make rational decisions as described by Kahneman in his book ``Thinking, Fast and Slow''. Kahneman identified two types of thinking process, one being intuitive and the other one as rational \cite{kahneman2011thinking}. The act of persuasion is linked with first type of thinking where no logical reasoning or calculations are required for decision making. %

%Yes!: 50 scientifically proven ways to be persuasive \cite{goldstein2008yes}
%Influence \cite{cialdini1987influence}

%\subsubsection{Emotional Intelligence and Affective Computing}

\subsection{Design of Persuasive Technology}

Technology has an impact on people's behavior. The challenge is how to design technology that can be accepted by user and also it amends their behavior for a positive change. The user acceptance is very important because the technology that is to be designed is not only to be used by users but it must change people's attitude and behavior towards some recognized goal. There are many concerns about the user acceptance of technology as changing the behavior by use of technology might encounter several ethical issues like privacy of sharing user personal data. 

In \cite{davis2009design}, the authors have discussed two methodological frameworks to enable persuasive technology development. One is \textit{Value Sensitive Design (VSD)} and second one is \textit{Participatory Design (PD)}.  In Value Sensitive Design, the designers are acknowledged with human value, their privacy and their autonomous behavior while in Participatory development, the designers themselves are equal stakeholders and are complete participants throughout the process of designing of technology. Both these frameworks can be applied to ensure a persuasive design for a new technology that can positively change people's attitude. 

 \begin{figure}[h]
\includegraphics[width=\linewidth]{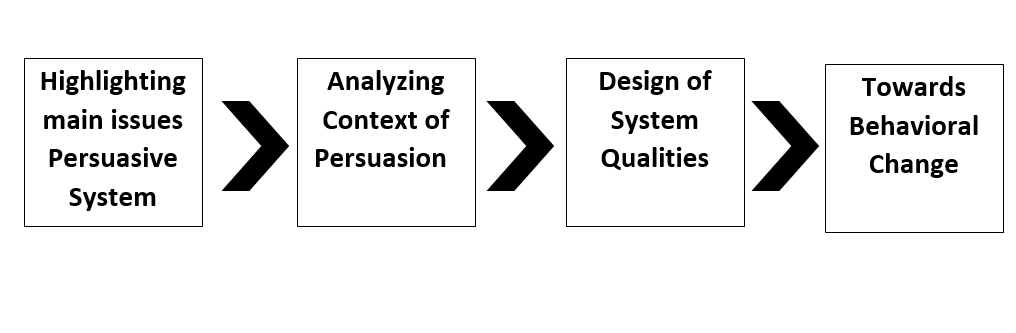}
\caption{Development Phases of Persuasive Systems}
\label{fig1:psd}
\end{figure}

Don Norman in his book ``Emotional Design'' \cite{norman2004emotional} explains more about designing of technology. It is argued that there is a relation between human cognition and emotions in the design of technology apart from logical design against usability of the product being designed for the users. The author claims that design that is attractive for users always ends up having more productive outcome as people are now relaxed towards the usage of technology.  Norman discussed three levels of design namely  \textit{visceral}, \textit{behavioral} and \textit{reflective}. \textit{Visceral} identifies the inward feelings of humans, the \textit{behavioral} stage identifies the feelings of people and about the usability of designed product and the \textit{reflective} design relates to the overall impact of designed product to the customers 

Another approach is ``Persuasive Systems Design'' (PSD), which represents a framework which describes what type of persuasive content is implemented and also it elaborates the contents and design of the final product as well \cite{oinas2009persuasive}. PSD is the process which describes how a persuasive system can be designed and also evaluates the system in context of its functionality for the users. The study explains the three major steps of developing of persuasive systems as shown in the figure \ref{fig1:psd}. The first stage is about understanding the key issues in designing of a persuasive system before its implementation. In the second stage, the context of persuasive system are to be examined. And in the final stage, the actual system is being implemented containing all the major aspects identified earlier.

To promote a healthier lifestyle for people who avoid exercising and neglect physical activities in their daily life, there are many solutions where persuasive technology can be used. One such study was conducted by researchers in which they designed the persuasive technology to help promoting physical activities. Using the existing design models, the authors have proposed a new model, based on previous work regarding persuasion to design system named, \textit{UbiFit Garden} system that aims to encourage physical activities among people using their smartphones \cite{consolvo2009theory}.

\section{Application of Persuasive Technology for Human Development}

In this section, we will elaborate different domains in which the persuasive technology is being used in everyday fields  such as health, education, physical activities, childcare, farming, energy usage, and fraud detection, etc. This section elaborates already implemented solutions to improve public interest using technology and persuasion.

\subsection{Health}
There is numerous work that is aimed to provide a healthier lifestyle for people and the role of information technology has played a vital part to give people healthier lifestyle by providing them with multiple applications for better health care. One of the most common approach is to provide people with choices, where they can choose a healthier option from multiple choices given to them and then feedback is provided on the choice made. In \cite{lee2011mining}, the authors have presented how they can promote the user to make choices which are in their own benefit, which is also the aim of behavioral economics where we are provided with certain choices and we will choose what is suitable for our own self. For the purpose of achieving their target, they conducted different experiments to study the effect of behavioral economics design and then tested to verify a positive change.

Another example where Information and Communication Technologies (ICTs) can be used is to provide women in India's slum areas with the information regarding child birth. The persuasion behind the defined approach is to provide speech-based information on mobile phones, to females so that it will influence the way by which most women face complications during the process of child birth \cite{ramachandran2010mobile}. One way to change behavior and adopt healthier lifestyle is that government imposes rules and regulations, for example enforcing high taxes on purchase of smoking products to avoid the illness caused by it, but the study of behavioral change tends to show that there can be other ways rather than just forcing to make choices for the well-being of people. One way to eradicate this problem is by giving incentives to people, For example in USA, people who were given \$750 per annum were three times more likely to quit smoking rather than those not given this incentive \cite{cahill2015incentives}. Another approach to handle the same issue of smoking cessation is by signing contracts in presence of other individuals which caused positive changes. 

Another major health issue is being overweight and obese (e.g., about six out of every ten adults are overweight\footnote{\url{http://www.gallup.com/poll/125741/six-overweight-obese.aspx}}) due to poor eating habits. Even though most people understand that eating fruits and fresh vegetables is beneficial for their health, still a large group of people fail to purchase fruits and vegetables when shopping. The use of behavioral insights and nudge theory can help direct people's eating behavior towards more healthier options. In an experiment, a university student designed the shopping trolley in such a way that it contains a yellow tape which divides the trolley in two sections, one for fruits and vegetables and other for remaining shopping items. Using this approach, there was a large increase in the amount of fruit and vegetables purchased by shoppers \cite{team2010applying}.

There are many other examples in which human behavior is changed using nudges, which is defined as making the person do some activity without forcing them to do that action. In health care and hygiene, the nudge theory can be implemented to provide better and healthier lifestyle. To encourage hygiene and avoid spillage in public toilets, a simple idea was to place a sticker of fly at appropriate place on men urinal so that it can become target for the person and it will ultimately avoid spillage on toilet floor. This approach, as it seems to be very simple and also humorous, tends to reduce the spillage to about 80\%. Another important strategy of nudge is providing less variety of food at cafeteria would make people eat less junk food and more healthier food which is limited in the menu of cafeteria \cite{oullier2010improving}. Some other methodologies for using nudges for better and healthier life (such as providing incentives, setting good defaults, salient and effect, social norms, subconsciousness and commitments) are described in \cite{blumenthal2012seeking}.

\subsection{Encouraging Physical Activity}
Today smartphones serve as an inlet to a person's mind and can be a potent tool for influencing a person. There are number of applications available that let the user interact with them to gain some benefit out of it. The persuasion theory can be applied to design such applications which promote physical activities in humans, one such example is \textit{Chick Clique}, which is a smartphone application that targets teenage girls who are cautious about gaining more weight and trying to loose weight using different techniques. This applications provides then with goals and information regarding their diet and also let teenage girls to socialize their activities to foster positive behavioral change \cite{toscos2006chick}.

A system involves designing using persuasive technology is \textit{Fit4life}, which is an iPhone application based on Persuasive System Design Model that is aimed to fight obesity in America (one of the main problems that afflict that region). It helps to determine calories in food eaten, an earphone which allows to directly communicate with the user, a heart rate monitor to check heart rate of the user and a sensor which is usually worn on toe to check current metabolic rate of user \cite{purpura2011fit4life}.

Human beings are keen to set goals in their daily life and indulge themselves in achieving those goals. The same idea is behind the application \textit{UbiFit}, which is a system that allows user to monitor their physical activities themselves using mobile phones. The main idea encouraging persuasion is setting of goals by user themselves and trying to achieve them daily as they are committed towards their own goals \cite{consolvo2009goal}.

\subsection{Education}

Persuasive technology can be applied to education sector as well to change the thinking of students at school level by changing the school setting positively. In \cite{mintz2012application}, the authors have developed an application that focus on instructional design in making learning for children more effective. The idea was to develop persuasive environment for children suffering from autism, a development disorder that appears about the age of three causing children to loose control over their body and behavior. A project called HANDS (Helping Autism Diagnosed young people Navigate and Develop Socially) was developed on mobile phone platform containing two main application interfaces, one web based toolkit which is used by teachers to design effective interfaces according to each child then they can link that intervention on child's mobile phone so that he can easily interact with it. The project proved to be very interactive for children suffering from Autism Spectrum Disorder (ASD) and helped to changes their attitude towards interaction.

Another technology that has been used extensively to affect human behavior is text messages. Text message reminders have been used in multiple domains to prompt individuals towards certain desirable actions. For example, Karlan et al. sent text message reminders to borrowers of certain microfinance banks in the Philippines. They found that when the text message included the name of the account officer's name, the message significantly, improved repayment. However, the authors also suggest taking into account the context, market characteristics and nature of institutions before designing ICT interventions \cite{fox2016behavioral}. In another study, Castleman et al. found that sending personalized text messages to low-income students reminding them to finish tasks led to a 5.7 percentage point increase in college enrollment from 66.4\% to 72.1\% \cite{castleman2015summer}. Similarly, the Behavioral Insights Team (BIT) sent a series of text messages at regular intervals. This led to a fall in dropout numbers at the end of the first semester by one third \cite{chande2015curbing}.

\subsection{Farming}
Farming is another domain where the idea of nudge is implemented to change the behavior of farmers who invest in purchasing of fertilizers to yield their crops. The idea behind this model is defined in \cite{duflo2011nudging}. The farmers usually invest in purchasing fertilizers when government provide subsidies on purchasing fertilizers but excess usage can damage the lands as well. So they provided limited time discounts on purchase of fertilizers during the time of harvest in the farms of Kenya which caused much profitable outcome for farmers.

\subsection{Child Care}
The behavior change in child care is another area where change in behavior is required so that it benefits the children. It explains the architecture as behavioral mapping in which there are several steps involved in this procedure starting from finding the problem and then creating behavioral map which involves initialization of the project, understanding the problem, understanding the clients and then diagnosing the solution \cite{gennetian2011behavioral}.

\subsection{Breaking Harmful Addictions}
Internet can be used to persuade people behavior and give them healthier life as it has power to motivate people to keep a better lifestyle. The Internet-based resources such as web-based applications help to make persuasive communication among people. Web based platforms can provide immediate feedback to people that help to add positive behavior, motivations and beliefs at different times. In this study \cite{lehto2011persuasive}, they have highlighted how different interventions using web based technologies can be used to persuade people to stop using alcohol and quit smoking habits. The most common methodologies include persuader which can be in the form of messages being delivered to user, self-monitoring include knowing one's self behavior to change for a positive intervention, simulation, personalization, tailoring and tunneling etc.

\subsection{Sustainability and Environmentalism}
Human beings use technology resources for better lifestyle but fully automated environment might result in frustration for the user as it might tend to be less user-friendly. On the other hand if technology has been built only using the behavioral approaches, that also has not shown very good results as it lacks many technical details in solution. To overcome the above mentioned scenario, there should be an excellent bonding between the technology and the user behavior. There are four major roles of technology defined, namely, 1) technology as \textit{intermediary}; 2) technology as an \textit{amplifier}; 3) technology as a \textit{determinant}; and 4) technology as a \textit{promoter} \cite{midden2008using}. In the paper \cite{midden2008using} the authors have investigated how technology can be used as promoter to make people aware of using natural resources in a sustainable way. They have shown that media has provided a lot of persuasion by showing signs, sayings directly to impact people to raise the environmental risk awareness. Also, they have defined how persuasive technology can be utilized to promote energy conservation by using artificial intelligent agents that play their role as a social actor to promote the positive behavioral change.

Using media resources to bring about behavioral change in people may lead to an unsuccessful outcome if they developer of Public Service Announcements (PSAs) omit the basic principles of media-oriented awareness. In \cite{bator2000application}, the authors have identified some special guidelines for the campaign designers to create public services messages. They highlighted that the designers should initially identify the target audience for their campaign and then test the reactions of audience using pilot messages. The designers would also be required to consider attitude persistence, memory and social norms to create messages and presentations that influence audience.

Social norms as discussed earlier have great influence on people behavior so the study was conducted, which resulted in claiming that global warming and climate changes can be reduced especially by studying certain area like conserving daily energy consumption at one's house and in offices, avoid pollution on roads by using public transports and supporting public policies set by government to reduce global warming. This was achieved by using descriptive social norms which are messages that influence people to behave under specific conditions. The use of such norms by higher authorities can effect people to change their behavior among bringing in climate changes \cite{griskevicius2008social}.

To protect the environment using persuasive approach which effects human behavior easily is to communicate using normative messages. The use of such norms  had a great effect on people behavior towards protection of natural resources and eco-friendly environment \cite{cialdini2003crafting}. The experiment was conducted by placing marked pieces of petrified wood along the path for park visitors and had descriptive messages, which tells what needs to be done in particular situation and second are the injunctive messages, which state what is approved or disapproved in particular situations. The result showed that descriptive messages resulted in more theft of petrified wood than injunctive norm messages. This approved that the persuasive approach to guide people using norm-based messages (descriptive or injunctive) effect the user behavior.

\subsection{Energy Use}
Saving energy is one of the key factors to provide clean environment which is pollution free. The government can play a vital role to make people behave more energy efficiently to provide them healthier life. In \cite{team2011behaviour} the authors have tested two different approaches to get behavioral insights from people and motivate them to ensure the saving of energy consumed. The first method is to how UK government has encouraged people to make their homes more green. This was done by giving locals incentives and also providing special discounts on the purchase of energy efficient household items at government level. The second approach is use of social norms to address people to be more energy savers. This was achieved by comparison of energy consumption with other users in the same vicinity. The government has also managed to limit itself to reduce the energy consumption of all departments by about 10\% within one year. The overall approach helped and motivated people to reduce energy consumption leading to a well-pleased lifestyle. Other researchers have found that making information easy to be accessed leads to people being more likely to act on it. For example, an email that contains a direct link to online energy information portal is much more likely to prompt people to engage with the information.

%\begin{figure}[h]
%\includegraphics[width=\linewidth]{fraud_1.png}
%\caption{Steps to reduce Fraud, Error and Debt}
%\end{figure}

\begin{table}[h!] 
\centering
\caption{Steps to reduce Fraud, Error and Debt, Adapted from \cite{team2012applying}}
\label{my-label}
\setlength{\tabcolsep}{18pt}
\renewcommand{\arraystretch}{1.5} 
%\begin{tabularx}{.8\linewidth}{|l|}
\begin{tabu} to 0.45 \textwidth { | X[l] |  }
\hline
 \textbf{Lessons from behavioral science ? seven steps to reduce fraud, error and debt} \\
The seven insights below are all based on evidence from behavioral science. They show that, by going with the grain of how people 	   behave, we can reduce the prevalence of fraud, error and debt. \\
\textbf{Insight 1. Make it easy:} Make it as straightforward as possible for people to pay tax or debts, for example by pre-populating a form with information already held.\\
\textbf{Insight 2. Highlight key messages:} Draw people's attention to important information or actions required of them, for example by highlighting them upfront in a letter. \\
\textbf{Insight 3. Use personal language:} Personalize language so that people understand why a message or process is relevant to them.\\
\textbf{Insight 4. Prompt honesty at key moments:} Ensure that people are prompted to be honest at key moments when filling in a form or answering questions.\\
\textbf{Insight 5. Tell people what others are doing:} Highlight the positive behaviour of others, for instance that `9 out of 10 people pay their tax on time'.\\
\textbf{Insight 6. Reward desired behaviour:} Actively incentives or reward behaviour that saves time or money.\\
\textbf{Insight 7. Highlight the risk and impact of dishonesty:} Emphasize the impact of fraud or late payment on public services, as well as the risk of audit and the consequences for those caught.\\
\hline
\end{tabu}
\end{table}

\begin{table}[h!] 
\centering
\caption{Trials to reduce Fraud, Error and Debt, Adapted from \cite{team2012applying}}
\label{my-label2}
\setlength{\tabcolsep}{18pt}
\renewcommand{\arraystretch}{1.5} 
%\begin{tabularx}{.8\linewidth}{|l|}
\begin{tabu} to 0.45 \textwidth { | X[l] |  }
\hline
\textbf{Test, learn, adapt ? eight trials to reduce fraud, error and debt}\\
\textbf{Trial 1. Using social norms:} investigates whether informing people that the vast majority of those in their area have already paid their tax can significantly boost payment rates.\\
\textbf{Trial 2. Highlighting key messages and norms:} examines whether we can increase tax compliance among doctors by simplifying the principal messages and actions required, as well as using social levers and norms.\\
\textbf{Trial 3. Using salient images:} investigates whether using images captured by the Driver and Vehicle Licensing Agency can help to reduce unnecessary repeat correspondence and encourage prompt payment of fines.\\
\textbf{Trial 4. Better presentation of information:} explores different ways of presenting information to discover which is most effective at encouraging the payment of debts.\\
\textbf{Trial 5. Personalizing text messages:} tests the impact of sending more personalized text messages on people's propensity to pay court-ordered fines.\\
\textbf{Trial 6. Prompting honesty:} examines whether simplifying key messages, emphasizing the consequences of fraud and getting people to sign forms upfront results in more honest declarations.\\
\textbf{Trial 7. Varying the tone of letters:} explores the effectiveness of different types of communication in encouraging plumbers to get their tax affairs up to date.\\
\textbf{Trial 8. Using beliefs about tax:} tests the effectiveness of different messages ? related to the fact that most people think that paying tax is the right thing to do ? on payment of tax debts by companies.\\
\hline
\end{tabu}
\end{table}

\subsection{Reducing Fraud, Error, and Debt}

Another important area where behavioral economics can play its part and change people mind to avoid frauds and negative attitude is reducing the frauds being committed. According to one study \cite{team2012applying}, the people in UK tend to stop frauds, errors and debt because of the strong social norm there, where people find it a very disgraceful to adopt one such habit. In this study, the effort of behavioral insight team to stop people from making fraud, errors and debts has been explained, also they have provided some trials that help to reduce the error, fraud and debt. The details of the two are shown in the tables \ref{my-label} and \ref{my-label2}.

\section{Case Study: Potential Applications of  Persuasive Technology for Development in Pakistan}

Pakistan is a country where there are many opportunities available to overcome the situation like poor health, high death rates and other social problems using persuasive methodology and using technology as well. One of the major areas to work using persuasive technology is in the domain of health, agriculture and changing social behavior of people to yield healthier and better lifestyle. Most of the countries including US have worked for their social problems using persuasive technology \cite{bator2000application} \cite{blumenthal2012seeking} \cite{purpura2011fit4life} \cite{toscos2006chick}.

\subsection{Improved Road Traffic Through Improved Driving Behavior}

According to World Health Rankings website, the rate of death caused in road accidents is about 20 per 100,000 people\footnote{\url{http://www.worldlifeexpectancy.com/cause-of-death/road-traffic-accidents/by-country/}}. The people who are injured in these situations is another count. The number of people that are directly or indirectly affected in the road accidents can be reduced by using the persuasive technology and behavioral science.

Smartphones, which have now become the central computer and communication devices in the lives of people around the world, have been shown in the literature to be very versatile in ICT for human development (ICTD) projects \cite{ali2016big} \cite{qadir2016crisis} \cite{abraham2006mobile}. Smartphones are especially well suited for capturing data about the user's behavior and for running persuasive technology applications that can be utilized to change people's behavior positively. A number of applications like \textit{Chick Clique}  \cite{toscos2006chick}, \textit{Fit4life} \cite{purpura2011fit4life} have used smartphones as primary source to promote healthy lifestyle among people. 

\begin{figure}[h]
\begin{center}
\includegraphics[width=.7\linewidth]{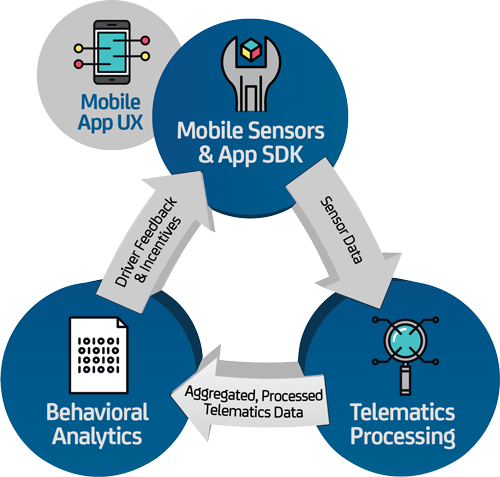}
\caption{Architecture of \textit{DriveWell} Application. The use of behavioural analytics and nudge theory can be used to improve driver's behaviour.}
\label{fig2:dw}
\end{center}
\end{figure}

In order to handle the issue of road accidents, different organizations including several insurance companies started to use persuasive technology as a tool to measure the behavior of drivers on road and then trying to change the attitude of drivers so that their driving is safe for environment as well as themselves. This system is known as Mobile Telematics System and major application which is developed by Cambridge Mobile Telemetics is \textit{DriveWell} \cite{malalur2013telematics}. The application architecture as described on their website\footnote{\url{https://www.cmtelematics.com/drivewell/}} is also shown in figure \ref{fig2:dw}. It describes the three layers of architecture for collection of sensory data, processing upon this received data and finally towards behavioral changes.

 \begin{figure*}[h]
 \begin{center}
\includegraphics[width=.65\linewidth]{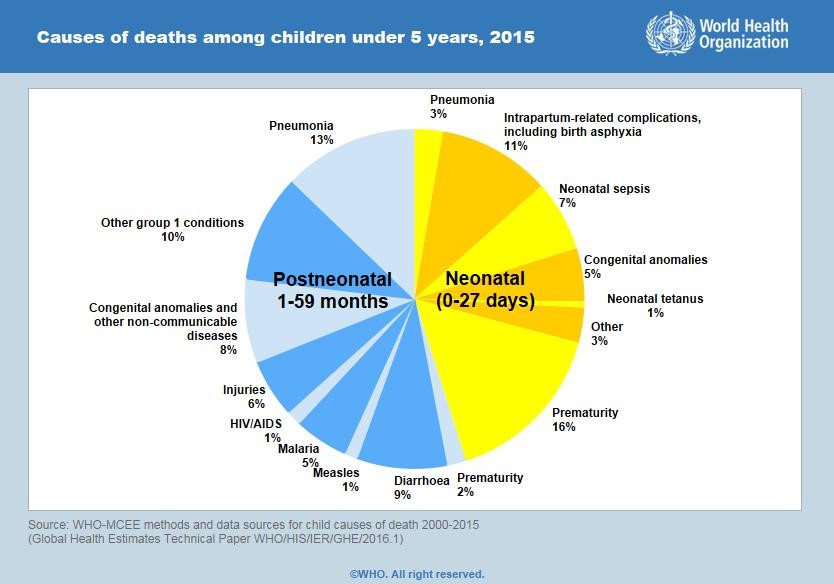}
\caption{Method and data sources for child causes of death 2000-2015. A large number of deaths caused are preventable through immunization, which motivates our nudge-based immunization education application.}
\label{fig4:himm}
\end{center}
\end{figure*}

Another application known as \textit{FLO}\footnote{\url{https://www.driveflo.com/}} is introduced which gives instant feedback to drivers on road to promote safe driving. The application aims to improve the driving pattern of drivers by using scoring mechanism, every trip of driver is recorded by application  and then positive scores are assigned for the trip according to safe driving, but if driving is judged as unsafe, then negative scores will be given which will affect driving of the person.

Based on these few above mentioned applications and to handle this global problems faced by people during driving, we aim to provide solution by building an Android based application for Pakistan, which can monitor the driving pattern of drivers on road using the built-in sensors of smartphone. The application features monitoring of driving actions that include over-speeding, tight cornering, unnecessary breaking, irregular acceleration, unnecessary lane switching and also phone usage during driving etc. The Android application will be run in background, accommodating all sensory data from the smartphone without diverting driver attention towards his phone aiming to provide the minimum distraction during driving. This application would follow Persuasive System Design and will utilize a nudge architecture to incline drivers towards responsible driving and away from rash driving. The users would be able to share their scores with their peers and ratings would be given so they would aim to get top of the leader board by better performance on driving.

%of giving rewards and benefits on safe driving and negative scoring on rash driving on roads as it will all be based on data collected from mobile phone. 

%The application would follow Persuasive System Design architecture to bring positive impact on every constant of driver which can have effect on his driving pattern. Every trip will monitor all the above mentioned parameters of driving and will accumulate the results in the form of positive and negative grading criteria, bench marking and color coding etc.

 %The data collected from sensors will be stored in a shared database (web application and mobile application) providing accessibility to users through this centralized system. The trip history would be stored and can be later viewed by driver and its care-taker. The care-taker would be the person that has social influence on driver like parents or owners of vehicle. This monitoring of person driving the vehicle is aimed to improve the driving patterns of drivers as they know that they are being monitored by their care takers. 

The aim of this application would be to provide safe journey for Pakistan as other application related to the same scenario are also built but they are operated by insurance companies to track driving of their clients, and not by common people themselves. This would be unique for country like Pakistan where number of road accidents is increasing.

%\subsection{Improved Vaccination Through Effective Messaging}

\subsection{Mobile Application Based Nudge framework For Education About Immunization}
After sub-Saharan Africa, Pakistan has the highest infant mortality rate. Extended Program of Immunization (EPI) in Pakistan is a government initiative which controls and monitors immunization in Pakistan. EPI suggests that 27\% of these deaths could be prevented through immunization. Disease like Pneumonia are the leading cause of infant mortality which can be easily averted through immunization.
 Most of these deaths are caused on prima facie due to neglect but are also are rooted in the disinformation disseminated through hearsay. Punjab Information Technology Board has made EVACS system to monitor the attendance of all vaccinators sent out in the field. However there is acknowledgment that the completion of vaccination courses has not improved. Parent used to forget their children's next vaccination date and for diseases like measles which have significant time period between two vaccination dates the dropout rates were high

 \begin{figure*}[h]
 \begin{center}
\includegraphics[width=.95\linewidth]{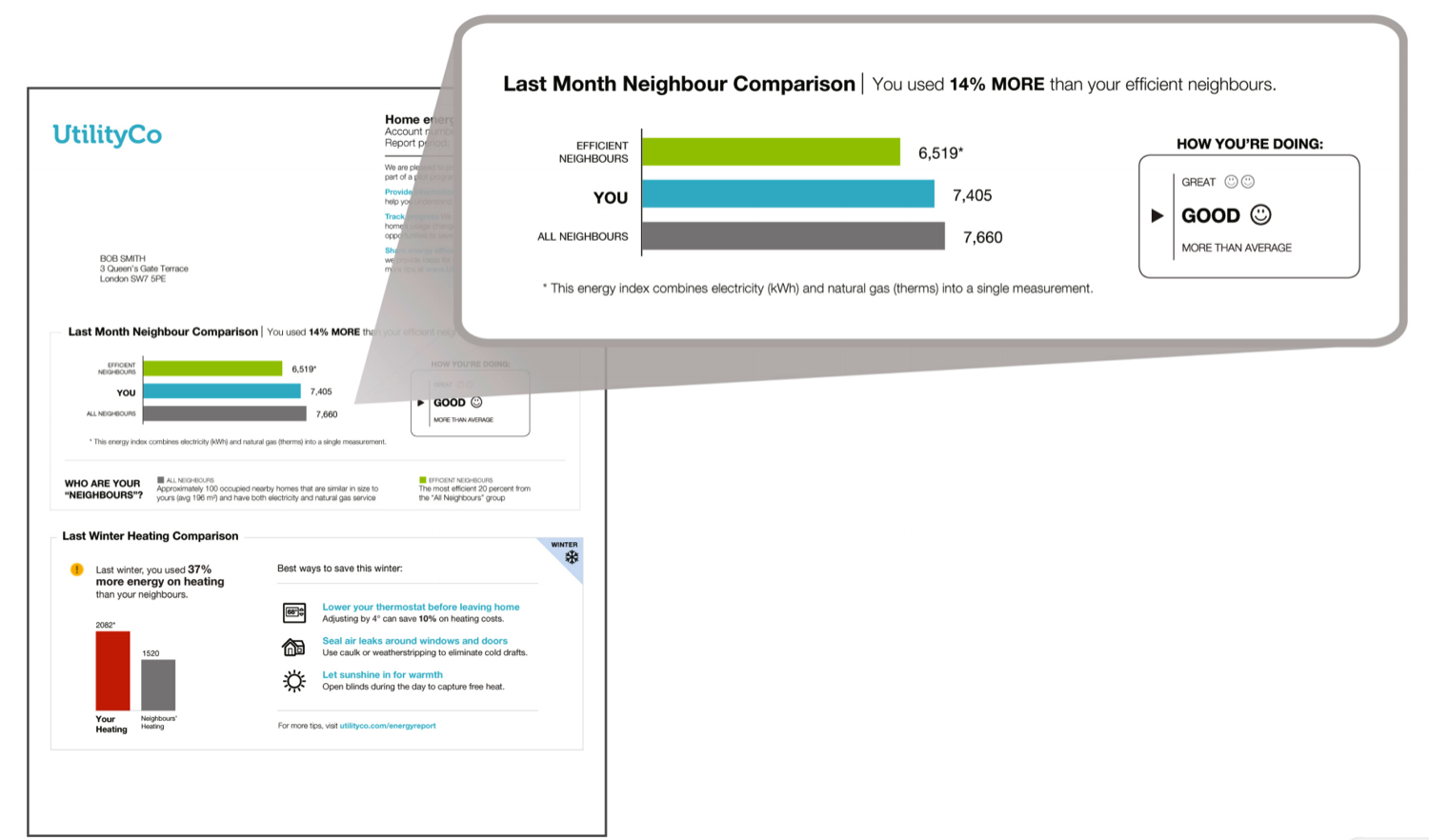}
\caption{A sample energy bill sent by company Opower to consumers that displays the energy consumed along with embedded nudges towards more energy-efficient future behavior. Figure source: \cite{team2011behaviour}}
\label{fig4:opower}
\end{center}
\end{figure*}

Large numbers of the premature births take place in rural areas and the death rate is high however in urban areas most of these births take place at tertiary care hospitals and the survival rate is generally better for these children. As shown in figure\ref{fig4:himm} there are numerous reasons for the cause of deaths of children and major part include most fatalities caused by premature deaths. Urbanization is growing in Pakistan and it is estimated (United Nations Population Division) that, by the end of 2025, nearly 50\% of Pakistan population would be living in cities. There are many reason for this, but the main reason is that living in urban cities increases the chances of better livelihood and access to better health and education services. Accompanied with this trend of urbanization is the growing number of urban smartphone users. Pakistan Start-up Report Estimates that there will be 110 smartphone users in Pakistan by the end of 2018 (most of which are users of affordable Android-based smartphones).

Our aim is to develop a smartphone application that not only stores the date of the next vaccination but also educates people about how to find the nearby Basic Health Unit (BHU) or other vaccination agencies and how to check the expiry of the vaccines. The most important aspect of the application is to nudge the parent towards vaccinating their children by message and videos from religious scholars and by streaming vaccination date to Facebook account of the users. The main focus of the application is to design the choice architecture that makes the users involved in improving the child and its mother's health. Framing of the messages in the application are also very important. The framing of the messages in the application would be such that the user is not alienated and should be such that persuasion techniques are subtle and motivate him to higher goals such as better education for his child. The other important factor in designing this application is that the social impact factor should also be taken into account and the users of the application could identify with his peers. The goal of the application thus is that through Libertarian Paternalism guide the individual towards sets of choices that lead to healthier life of his dependents.

\subsection{Limiting Energy Consumption}

Pakistan is growing at a fast pace and has a rapidly rising population. The rise in population is stressing the infrastructural capacity of Pakistan, particularly in the area of electricity generation. Currently, the demand of electricity has far outstripped the generational capacity of current infrastructure, resulting in an average shortfall of 4,000 Megawatts. Due to this massive energy deficit, the everyday life of Pakistani citizens is significantly affected and it is common for urban citizens to face up to 6--12 hours of \textit{load shedding}\footnote{Load shedding refers to the action of shedding excessive load above generating capacity by interrupting electricity supply to consumers} and rural areas suffer even more (18--21 hours of load shedding). While the Government of Pakistan has initiated many projects in recent times to improve the generational infrastructure, due to the concomitant increase in the population and therefore the user demand, it is anticipated that the practice of load shedding may not be completely eliminated unless consumer behavior is also transformed.

The use of persuasive technology and behavioral insights can be profitable here as has been shown in previous efforts in literature and practice (particularly, in the UK and US) \cite{team2011behaviour}. We can reduce the energy consumption substantially by developing new social norms in which energy saving is encouraged. Various techniques of influence and persuasion can be utilized (e.g., comparison with other users in the vicinity). The UK government has developed a Behavioral Insights Team (also known as the Nudge Unit)  \cite{team2011behaviour} that has been successful in encouraging people to make their homes more green. This was done by giving people incentives and also providing special discounts on the purchase of energy efficient household items at government level. 

As another example, A US-based company Opower\footnote{\url{http://www.opower.com}} sends electricity and gas consumption reports to consumers by comparing their usage with other costumers of similar household usage. A snapshot of the power bill sent by Opower is shared in Figure \ref{fig4:opower}. This figure encapsulates an embedded nudge towards more energy efficient behavior since potential benefits are clearly shown and a more energy-efficient social norm is emphasized by comparing the subscriber to other similar subscribers. Users are also provided energy conservation tips on their bills and this practice had a positive result and people started to restrict their consumption by 2\$ \cite{allcott2010behavior}. The policy makers identified three major policies to help to achieve the goal of saving energy. First is the financial help from government, Second is providing users with incentives against their consumption of gas and electricity and third is that government should provide information disclosure regarding the departments to promote more energy savings. The same ideas can also be used to curb energy consumption and can help mitigate the hard to resolve energy crisis.

\section{Opportunities and Pitfalls}

There are many issues regarding the use of persuasive technology as a tool to modify human behavior. Few such pitfalls and opportunities are described in this section. 

\subsection{Ethics Issue:}
Nudging as it applied to make changes in behavior of peoples might not prove to be ethical. The government rules imposed to people might have negative effective on people and they lack choices. for example, in case of giving defaults (a type of nudge), there are no or limited choices for individuals to choose from so it proves to be false choice for people. The nudge or changes in choice architecture may not be acceptable on ethical grounds as they tend to violate on three major grounds which are welfare, autonomy and dignity of people \cite{sunstein2014ethics}. Other examples that might affect people include Reminders, a nudge, but still it is like a warning, so it must be clear that nudge should not impose incentives which are merely wrong choices for people rather it must provide choices which are acceptable on all ethical grounds as well.
The use of technology can be beneficial but if that is forced on people as part of nudging, it becomes very complicated for people to adopt to such technology. The usage of technology would let us decide how to adopt that technology as an example The most banal example from Bruno Latour, in reference to the American National Rifle Association's assertion that guns do not kill people, but that it is people who kill people \cite{verbeek2011moralizing}.

\subsection{Privacy Issues:}
The nudging can also lead to privacy issues as the data collected from mobile phones like location to identify people's position and who is nearby them, the patterns of purchasing on application stores enables the interest of people, the browsing patterns of people identify what type of contents the user is interested in, Social networking websites collect data of users to know their interests and that leads to appropriate advertisement contents being generated. All this private information can be used by government and private agencies to keep record of consumers when people do not know that they are already sharing this sacred information over the Internet. This leads to privacy violation under the umbrella of nudge \cite{schneier2015data}.

\subsection{Algorithmic Harms:}
Algorithms have been designed to provide facility to people so that their complex problems can be solved within no time but still such type of approach when applied in real world environment could be a loss of people's interest. As algorithms written by Facebook and Google provide user's confidential data to these companies causing privacy issues. A real world example is that there are many companies which are using the algorithms for hiring or maybe firing individuals. The decisions which need to be taken by humans solely depend upon applied algorithms which might violate many of the government regulations as well \cite{tufekci2015algorithmic}. A book written by Cathy O'Neil titled ``Weapons of Math Destruction'' \cite{o2016weapons} also aims to shed light on the above mentioned dilemma. The author has worked to provide people with enough knowledge that the use of algorithmic approaches has negative effects on human behavior as many wrong decisions could be the outcome of these algorithms, however these algorithms, which are merely mathematical models, should be much fair and they should judge without bias but the reality is different as they lack many emotional and psychological factors in making decisions. 

\subsection{Propaganda and Exploitation:}
Massive production of data has enabled huge data pools and these technologies have contributed in massive production of big data and computational practices that enable to produce high quality results by measuring the shifts and trends using tools and technologies. Computational politics is one of the area which involves applying the computational models to huge data sets which are derived from online and offline sources of data to conduct political activities like electing, opposing a candidate, or a government policy which is an act of persuasion for all individuals. In \cite{tufekci2014engineering}, the authors have provided how computational politics combined with big data and focus on the power of big data to help local public. In another study \cite{helbing2017will}, it is conducted that how big data, which is increasingly growing by massive production of data by users, can be combined with artificial intelligence, a modernized technique in computing in which the aim is to make everything much intelligent by inducing the power to decide the matters on behalf of machine itself. These latest trends in behavioral science and technology are reshaping our society and taking control in making decisions related to public affairs etc.

\section{Conclusions} 

In this paper, we have given a broad overview of how persuasive technology can be used to facilitate human development, particularly, in the setting of the developing world. After providing a self-contained introduction to the background of persuasive technology, and highlighting common applications of persuasive technology in literature, we discuss the particular case of Pakistan, a typical developing country in which there is a need to change the behaviour of people in many sectors of daily life. We present an insight into the unique problems faced by Pakistan and present three applications (safe driving, pro-vaccination nudging, and responsible energy consumption) of how persuasive technologies can help improve human lives in Pakistan. Finally, we note some pitfalls that may accompany the use of persuasive technology for human development.

\newpage

\bibliographystyle{plain}
\bibliography{persuasive}

\end{document}